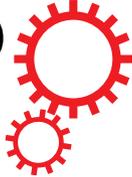

# OPEN

# Electron waiting times in hybrid junctions with topological superconductors

Shuo Mi[1,2], Pablo Burset[1] & Christian Flindt[1]

We investigate the waiting time distributions (WTDs) of superconducting hybrid junctions, considering both conventional and topologically nontrivial superconductors hosting Majorana bound states at their edges. To this end, we employ a scattering matrix formalism that allows us to evaluate the waiting times between the transmissions and reflections of electrons or holes. Specifically, we analyze normal-metal–superconductor (NIS) junctions and NISIN junctions, where Cooper pairs are spatially split into different leads. The distribution of waiting times is sensitive to the simultaneous reflection of electrons and holes, which is enhanced by the zero-energy state in topological superconductors. For the NISIN junctions, the WTDs of trivial superconductors feature a sharp dependence on the applied voltage, while for topological ones they are mostly independent of it. This particular voltage dependence is again connected to the presence of topological edge states, showing that WTDs are a promising tool for identifying topological superconductivity.

Topological superconductivity is an exotic state of quantum matter characterized by the emergence of symmetry-protected gapless edge states[1]. Such excitations are the condensed matter realization of Majorana fermions[2–4]. Due to their topological protection against disorder and their non-Abelian statistics, Majorana modes are a prominent building block for topological quantum computers[5,6]. The search for topological superconductors has thus become the focus of an intense research activity. Topological superconductivity naturally arises in a few superconducting materials[7], however, most commonly, it is a result of careful material engineering. In one approach, an effective *p*-wave pairing is induced by the proximity-effect from conventional *s*-wave superconductors on materials with strong spin-orbit coupling[2–4]. The *p*-wave pairing can then be controlled by an external field or magnetic impurities[8–15]. To detect Majorana modes, one can measure the tunnel conductance, which scans the surface density of states of the superconductor[16,17]. Zero-bias conductance peaks are related to the presence of surface states and thus provide signatures of topological and other types of unconventional superconductivity[18,19]; cf. Fig. 1(a). However, conductance and shot noise measurements are sensitive to impurity scattering and temperature broadening[20–22] and require junctions in the tunneling regime to provide unambiguous signatures of topological superconductivity.

An alternative to conventional current measurements is provided by the recent progress in the real-time detection of single electrons in nano-scale systems. Electron counting techniques have by now reached a level of sophistication where single charges can be manipulated and detected with high precision[23–26], opening a wide range of possibilities for exploring the statistics of electron transport[27–29]. One may for example investigate the waiting time between subsequent tunneling events[30]. The waiting time distributions (WTDs) provide information about the internal dynamics of a mesoscopic system which is useful for systems with localized states[31–34]. Theories of WTDs for mesoscopic conductors have recently been developed[35–37] and used to investigate the regularity of dynamic single-electron emitters[38–44]. Experimentally, it is possible to detect single Andreev processes at normal-metal–superconductor interfaces[45,46]. For energies below the superconducting gap, electrons in the normal-metal are converted into Cooper pairs in the superconductor, leaving a hole behind, see Fig. 1(b). Thus, the Andreev-reflected hole in the normal-lead is directly connected to the transfer of a Cooper pair into the superconductor. In turn, the WTD of reflected holes is equivalent to that of the transferred Cooper pairs, and it may unravel the internal dynamics of a superconducting hybrid[47] or reveal the presence of entangled electron-hole pairs[48,49]. For topological superconductors, a spin-sensitive single-electron detector would observe a distinct signal from a Majorana mode at very low voltages[50].

[1]Department of Applied Physics, Aalto University, 00076, Aalto, Finland. [2]Univ. Grenoble Alpes, CEA, INAC-Pheliqs, 38000, Grenoble, France. Shuo Mi and Pablo Burset contributed equally. Correspondence and requests for materials should be addressed to S.M. (email: shuo.mi@univ-grenoble-alpes.fr)

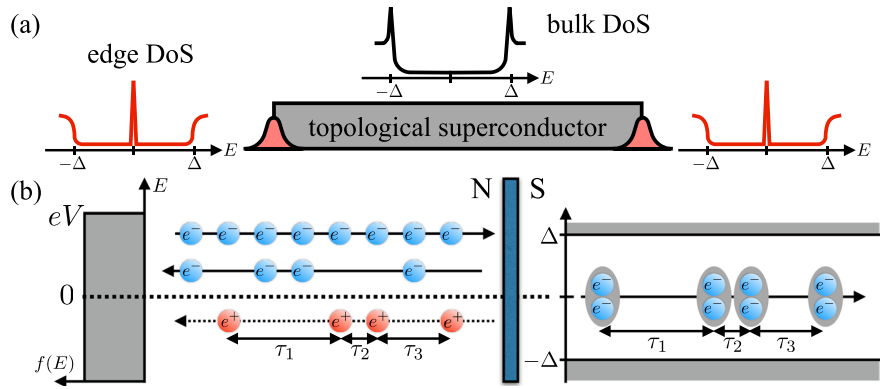

**Figure 1.** Electron waiting times of superconducting hybrids. (**a**) One-dimensional topological superconductor with two edge states: the bulk density of states (DoS) shows a conventional gap (black line), while the edge states appear as zero-energy peaks (red lines). (**b**) Schematics of an NIS junction, where an applied dc voltage drives electrons from the normal-state electrode (N) into a trivial or topological superconductor (S). Electrons with excitation energy smaller than the superconducting gap $\Delta$ are Andreev reflected and a Cooper pair is transmitted into the superconductor. Incident electrons can also be normal-reflected at the interface. The time between the reflection of two successive holes and the injection of two Cooper pairs is the same.

In this work, we identify the presence of edge states in topological superconductors using the distribution of electron waiting times. We describe the coherent microscopic processes that take place at normal-metal–superconductor (NIS) junctions with conventional or topological superconductors and include the effects of the transmission amplitudes and the applied voltage. Specifically, we demonstrate that the WTDs of Andreev processes are sensitive to the presence of topological edge states even under conditions where conductance and noise measurements would be ambiguous, i.e., in the presence of impurities or with high transmission amplitudes. While conductance spectroscopy requires tunnel junctions and fine-tunning of the bias voltage for subgap values, the distribution of electron waiting times can identify topological superconductors at junctions with arbitrary transparency and for fixed voltages. In an NISIN setup with grounded superconductor, we show that the WTD is sensitive to the number of resonances inside the transport window. For trivial superconductors, the WTDs change abruptly from a low-transmission profile for subgap voltages into an oscillatory one that reflects the transfer of electrons for voltages above the superconducting gap. The WTDs of topological superconductors, by contrast, are mainly determined by the zero-energy state and do not change over a wide range of voltages.

This rest of the paper is now organized as follows. In Sec. 2, we describe the microscopic scattering theory of normal-metal–superconductor junctions. In Sec. 3, we discuss the theory of WTDs for coherent scatterers. In Sec. 4, we analyze the WTDs of NIS junctions with one normal-state lead and one superconductor lead. In Sec. 5, we turn to NISIN junctions, where the superconductor is sandwiched between two normal-state leads, so that Cooper pairs can be spatially split into different leads. Finally, Sec. 6 contains our concluding remarks and an outlook on prospects for future work.

## Superconducting Junctions

We are interested in the electronic transport in hybrid junctions consisting of superconducting and normal-state regions. The low-energy excitations of a superconductor are well-described by the Hamiltonian[51]

$$H(\mathbf{k}) = \begin{pmatrix} [\varepsilon(\mathbf{k}) - \mu]\hat{\sigma}_0 & \hat{\Delta}(\mathbf{k}) \\ -\hat{\Delta}^*(-\mathbf{k}) & [\mu - \varepsilon(-\mathbf{k})]\hat{\sigma}_0 \end{pmatrix}, \quad (1)$$

where $\mu$ is the chemical potential and $\hat{\sigma}_{0,1,2,3}$ are the Pauli matrices in spin space. Corresponding to our choice of basis, the quasi-particle state has the form

$$\Psi(\mathbf{k}) = [u_\uparrow(\mathbf{k}), u_\downarrow(\mathbf{k}), v_\uparrow(\mathbf{k}), v_\downarrow(\mathbf{k})]^T, \quad (2)$$

where $u_\sigma(\mathbf{k})$ and $v_\sigma(\mathbf{k})$ are the electron- and hole-like components with spin $\sigma = \uparrow, \downarrow$ and wave vector $\mathbf{k}$. We focus on one-dimensional systems extending along the $x$-direction and electronic excitations near the Fermi surface with the dispersion relation

$$\varepsilon(\mathbf{k}) = \frac{\hbar^2 k_x^2}{2m}, \quad (3)$$

where $m$ is the effective mass. For a singlet $s$-wave superconductor, the pairing potential has the simple form

$$\hat{\Delta}(\mathbf{k}) = i\Delta e^{i\phi}\hat{\sigma}_2, \qquad (4)$$

where the amplitude $\Delta \geq 0$ is independent of $\mathbf{k}$ and $\phi$ is the superconducting phase. A triplet $p$-wave superconductor, by contrast, has a pairing potential that is an odd function of the wave vector. Below, we choose the spin quantization axis to lie along the $z$-direction and to be parallel to the spin polarization of the triplet state[18,52]. The resulting pairing potential is then

$$\hat{\Delta}(\mathbf{k}) = \Delta(k_x/|\mathbf{k}|)e^{i\phi}\hat{\sigma}_1, \qquad (5)$$

which is an odd function of $k_x$.

We now consider the Bogoliubov-de Gennes equations

$$H(\mathbf{k})\Psi(\mathbf{k}) = E\Psi(\mathbf{k}), \qquad (6)$$

where $E$ is the excitation energy measured from the Fermi level. To proceed, we note that for both singlet and triplet superconductors, we can decouple the two independent spin channels of the Hamiltonian in Eq. (1). For each spin channel, we then find solutions of the form

$$\psi_\sigma(x) = \begin{pmatrix} u_\sigma(x) \\ v_\sigma(x) \end{pmatrix} = \sum_{\alpha=\pm}\left\{ a_{\sigma\alpha}\begin{pmatrix} u_0 \\ \eta_{\sigma\alpha}^* v_0 \end{pmatrix}e^{\alpha i k_1 x} + b_{\sigma\alpha}\begin{pmatrix} \eta_{\sigma\alpha} v_0 \\ u_0 \end{pmatrix}e^{\alpha i k_2 x} \right\}. \qquad (7)$$

Here, the two wave vectors are given by

$$\hbar k_{1,2} = \sqrt{2m(\mu \pm \sqrt{E^2 - \Delta^2})}, \qquad (8)$$

and the ratio of the hole and electron amplitudes reads

$$\frac{v_0}{u_0} = \frac{\Delta e^{i\phi}}{E - \sqrt{E^2 - \Delta^2}}. \qquad (9)$$

Above, we have defined $\eta_{\sigma\alpha} = \pm 1$ depending on the spin $\sigma$ for singlet superconductors and on $\alpha$ for triplet superconductors. Thus, for a specific geometry, we need to find the coefficients $v_0$, $a_{\sigma\alpha}$ and $b_{\sigma\alpha}$, so that the boundary conditions are fulfilled and the state is normalized. We note here that a superposition of the $s$- and $p$-wave pairings from Eq. (4) and Eq. (5) will also lead to a decoupling of the two spin channels of the Hamiltonian. Following ref.[53], Eqs (7–9) can thus be easily extended to account for this type of pairing.

Throughout this work, we will be interested in transport processes taking place in hybrid systems involving normal-state regions ($\Delta = 0$) and superconducting regions ($\Delta \neq 0$). As an example we first describe an NIS junction consisting of a normal-state (N) region connected via an insulating barrier (I) to a superconductor (S). The insulator is described by the potential barrier

$$V(x) = Z\frac{\hbar^2 k_F}{2m}\delta(x) \qquad (10)$$

positioned (at $x = 0$) between the normal-state region ($x < 0$) and the superconductor ($x > 0$). Here, $k_F = \sqrt{2m\mu}/\hbar$ is the Fermi wave vector, and $Z$ is the dimensionless strength of the barrier[54]. We assume that the amplitude of the pairing potential changes abruptly from $\Delta$ in S to zero in N. This assumption is valid if the Fermi wavelength in S is much smaller than the proximity-induced coherence length $\xi = \hbar v_F/\Delta$. The corresponding boundary conditions then read

$$\psi_\sigma(0^-) = \psi_\sigma(0^+), \quad k_F Z\psi_\sigma(0) = \partial_x \psi_\sigma(x)|_{x=0^+} - \partial_x \psi_\sigma(x)|_{x=0^-}. \qquad (11)$$

We consider electrons being injected from the normal-metal and correspondingly set $a_{\sigma+} = 1$, $a_{\sigma-} = r_{ee}^\sigma$, $b_{\sigma+} = r_{eh}^\sigma$, and $b_{\sigma-} = 0$ in Eq. (7) for $x < 0$. Here, $r_{ee}^\sigma$ and $r_{eh}^\sigma$ are the amplitudes for incoming electrons to be reflected by the superconductor, coming back as an electron (normal reflection) or a hole (Andreev reflection). Imposing the boundary conditions above and requiring the state to be normalized, the reflection amplitudes become

$$r_{ee}^\sigma = -\frac{Z(Z+i)(u_0^2 - v_0^2\eta_{\sigma+}\eta_{\sigma-})}{u_0^2 + Z^2(u_0^2 - v_0^2\eta_{\sigma+}\eta_{\sigma-})}, \quad r_{eh}^\sigma = \frac{u_0 v_0 \eta_{\sigma+}}{u_0^2 + Z^2(u_0^2 - v_0^2\eta_{\sigma+}\eta_{\sigma-})}. \qquad (12)$$

For singlet superconductors, the amplitudes are the same for both spin directions, up to an irrelevant sign in the Andreev reflection amplitude $r_{eh}^\sigma$. For the triplet superconductors, the amplitudes for the two spin channels are identical. The difference between singlet and triplet superconductors comes from the product $\eta_{\sigma+}\eta_{\sigma-}$ in the denominator of the amplitudes. For singlet superconductors, we have $\eta_{\sigma+}\eta_{\sigma-} = 1$, while for triplet superconductors, we get $\eta_{\sigma+}\eta_{\sigma-} = -1$, leading to the formation of a zero-energy Andreev bound state[18,55–57].

The procedure above can be used to find the transmission and reflection amplitudes of more complicated systems, for instance with two normal leads coupled to a superconductor[58–62]. In the NISIN junction, a superconductor of width $d_S$ is coupled via insulating barriers to two normal-metal leads. For the left insulating barrier, we impose the boundary conditions in Eq. (11), substituting $Z$ by $Z_L$. For the right barrier (at $x = d_S$), we similarly have

$$\psi_\sigma(d_S - 0^+) = \psi_\sigma(d_S + 0^+), \quad k_F Z_R \psi_\sigma(d_S) = \partial_x \psi_\sigma(x)|_{d_S+0^+} - \partial_x \psi_\sigma(x)|_{d_S-0^+}. \tag{13}$$

Having formulated the boundary conditions, we can then evaluate the scattering properties of the NISIN structure.

### Electron Waiting Times

We are interested in quantum transport through superconducting hybrid junctions, specifically in the waiting time between quasi-particles leaving the superconducting region[48,50]. For example, one may consider the waiting time between an electron with spin-up leaving the superconductor and the next hole with spin-down leaving the superconductor. The waiting time is a fluctuating quantity, which must be described by a probability distribution. The waiting time distribution (WTD) is the conditional probability density of detecting a particle of type $\beta$ at time $t_\beta^e$, given that the last detection of a particle of type $\alpha$ occurred at the earlier time $t_\alpha^s$. Here, the types $\alpha$ and $\beta$ may refer to the out-going channel, the spin of the particle, and the particle being an electron or a hole. The WTD is denoted as $\mathcal{W}_{\alpha \to \beta}(t_\alpha^s, t_\beta^e)$. For the systems considered here with no explicit time dependence, the WTD is a function only of the time difference, such that $\mathcal{W}_{\alpha \to \beta}(t_\alpha^s, t_\beta^e) = \mathcal{W}_{\alpha \to \beta}(\tau)$ with $\tau = t_\beta^e - t_\alpha^s$ [35].

To evaluate the WTD, we proceed as in ref.[37] and express the WTD as time-derivatives of the idle time probability. The idle time probability $\Pi(t_s^\alpha, t_e^\alpha)$ is the probability that no particles of type $\alpha$ are detected in the time interval $[t_s^\alpha, t_e^\alpha]$ by a detector at position $x_\alpha$. The idle time probability can be a function of several different particle types and associated time intervals. The WTD can be related to the idle time probability by realizing that time-derivatives correspond to detection events. Specifically, the distribution of waiting times between particles of type $\alpha$ and particles of type $\beta$ can be expressed as[37]

$$I_\alpha \mathcal{W}_{\alpha \to \beta}(\tau) = -\partial_{t_\beta^e}\partial_{t_\alpha^s}\Pi(t_\alpha^s, t_\alpha^e; t_\beta^s, t_\beta^e)\big|_{t_\alpha^s, t_\beta^s, t_\beta^e, \to 0}^{t_\beta^e \to \tau}, \tag{14}$$

where $I_\alpha$ is the average particle current of type $\alpha$ particles, and the minus sign comes together with the derivative with respect to the starting time $t_\alpha^s$. In addition, after having performed the derivatives, we set the starting times to zero, i.e., $t_\alpha^s = t_\beta^s = 0$, while for the end times we set $t_\alpha^e = 0$ and $t_\beta^e = \tau$. The waiting time is then measured from the time when a particle of type $\alpha$ is detected until the later time when a particle of type $\beta$ is detected. During this waiting time, additional particles of type $\alpha$ may be detected, but not of type $\beta$.

The idle time probability can be evaluated using scattering theory, leading to the determinant formula[37]

$$\Pi(\{t_\gamma^s, t_\gamma^e\}) = \det[\mathbb{1} - \mathcal{Q}(\{t_\gamma^s, t_\gamma^e\})], \tag{15}$$

where the set $\{t_\gamma^s, t_\gamma^e\}$ corresponds to all relevant particles and associated time intervals. The hermitian operator $\mathcal{Q}(\{t_\gamma^s, t_\gamma^e\})$ is a matrix in the combined energy and particle type representation. It has the block form

$$[\mathcal{Q}]_{EE'} = \mathcal{S}^\dagger(E)\,\mathcal{K}\,(E - E')\,\mathcal{S}(E'), \tag{16}$$

having omitted the time arguments. The scattering matrix $\mathcal{S}(E)$ and the kernel $\mathcal{K}(E)$ are matrices in the space of particle types. The kernel is the block diagonal matrix

$$\mathcal{K}(\{t_\gamma^s, t_\gamma^e\}); E) = \bigoplus_\gamma K(t_\gamma^s, t_\gamma^e; E) \tag{17}$$

given by the direct sum of kernels

$$K(t_\gamma^s, t_\gamma^e; E) = \frac{\kappa}{\pi E} e^{-i\frac{E}{2}\left(t_\gamma^e + t_\gamma^s + \frac{2x_\gamma}{\hbar v_F}\right)} \sin(E(t_\gamma^e - t_\gamma^s)/2) \tag{18}$$

corresponding to each particle of type $\gamma$ with a detector at position $x_\gamma$. We work close to the Fermi level, where the dispersion relation $E = \hbar v_F k$ is approximately linear and all quasi-particles propagate with the Fermi velocity $v_F$. To implement the matrix in Eq. (16), we discretize the transport window $[E_F, E_F + eV]$ in intervals of width $\kappa = eV/\mathcal{N}$, where $\mathcal{N}$ is the total number of intervals. The width $\kappa$ enters in Eq. (18), and we always consider the limit $\mathcal{N} \to \infty$, for which the transport is stationary.

By combining Eqs (14) and (15), we now find

$$I_\alpha \mathcal{W}_{\alpha \to \beta}(\tau) = \frac{1}{\Pi}\widetilde{\mathcal{F}}_\alpha \mathcal{F}_\beta + \Pi \text{Tr}\left\{\mathcal{G}\frac{\partial \mathcal{Q}}{\partial t_\beta^e}\mathcal{G}\frac{\partial \mathcal{Q}}{\partial t_\alpha^s}\right\}\bigg|_{t_\alpha^s, t_\beta^s, t_\beta^e, \to 0}^{t_\beta^e \to \tau}, \tag{19}$$

having made repeatedly use of Jacobi's formula for derivatives of determinants, and we have defined

$$\mathcal{G} = (\mathbb{1} - \mathcal{Q})^{-1}. \tag{20}$$

In addition, the first-passage time distributions read

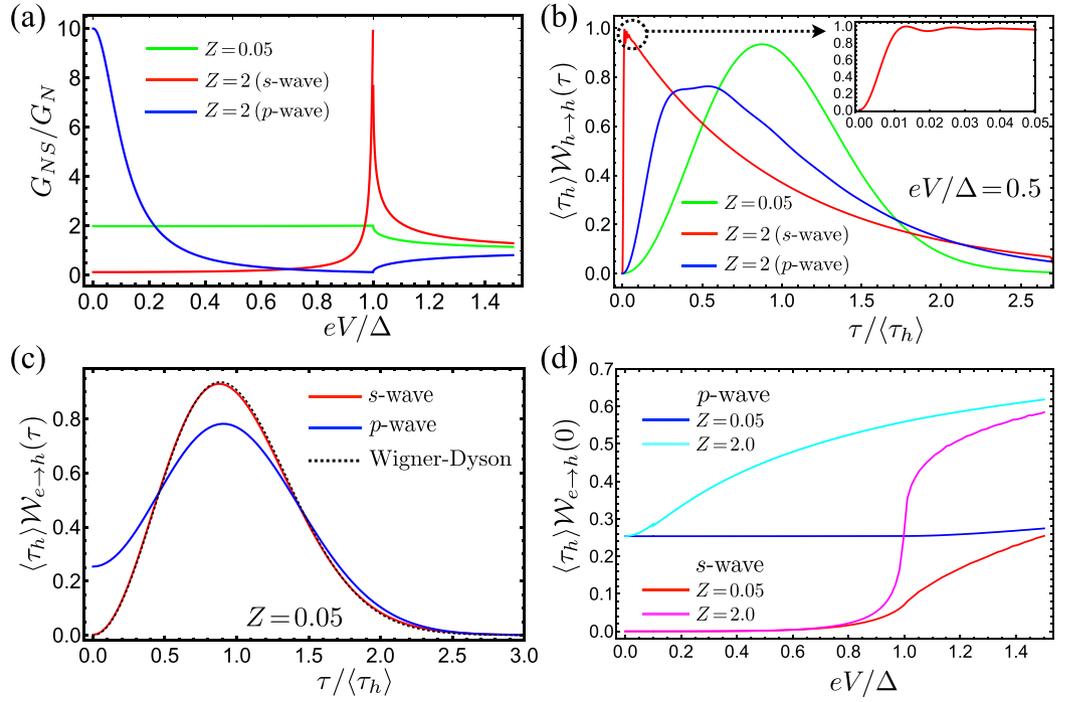

**Figure 2.** Electron waiting times of NIS junctions. (**a**) For transparent ($Z \simeq 0$) NIS junctions, the differential conductance for *s*- and *p*-wave superconductors is the same (green line with $Z = 0.05$). For tunnel junctions, the conductance features a gapped profile for *s*-wave pairing (red line) and displays a zero-bias peak for *p*-wave (blue line). (**b**) WTD between reflected holes at fixed voltage $eV = \Delta/2$ for the same parameters as in (**a**). Inset: short-time WTD for *s*-wave pairing. (**c**) Distribution of waiting times between electrons and holes for a highly transparent junction ($Z = 0.05$). The black dotted line shows the Wigner-Dyson distribution. (**d**) Distribution of waiting times between reflected electrons and holes at zero waiting time. Here, as a function of the applied voltage.

$$\widetilde{\mathcal{F}}_\alpha(\tau) = \partial_{t_\alpha^s}\Pi\big|_{t_\alpha^s,t_\beta^s,t_\beta^e,\to 0}^{t_\beta^e \to \tau} = -\Pi \operatorname{Tr}\left\{\mathcal{G}\frac{\partial \mathcal{Q}}{\partial t_\alpha^s}\right\}\bigg|_{t_\alpha^s,t_\beta^s,t_\beta^e,\to 0}^{t_\beta^e \to \tau} \tag{21}$$

and

$$\mathcal{F}_\beta(\tau) = -\partial_{t_\beta^e}\Pi\big|_{t_\alpha^s,t_\beta^s,t_\beta^e,\to 0}^{t_\beta^e \to \tau} = \Pi \operatorname{Tr}\left\{\mathcal{G}\frac{\partial \mathcal{Q}}{\partial t_\beta^e}\right\}\bigg|_{t_\alpha^s,t_\beta^s,t_\beta^e,\to 0}^{t_\beta^e \to \tau}. \tag{22}$$

The first distribution, Eq. (21), is the conditional probability density that no particles of type $\beta$ are detected in the time span $[0, \tau]$, given that a particle of type $\alpha$ was detected at the initial time $t = 0$. The second distribution, Eq. (22), concerns the time $\tau$ we have to wait until a particle of type $\alpha$ is detected, given that we start the clock at time $t = 0$. Finally, for evaluating Eq. (19) we note that the average particle current of type $\alpha$ particles can be expressed as $I_\alpha = \mathcal{F}_\alpha(0)$. In combination, Eqs (15–22) allow us to evaluate the distributions of waiting times for the superconducting systems that we consider in the following.

### NIS Junctions

We first consider an NIS junction. The differential conductance at zero temperature is given by the well-known expression[54]

$$G_{NS} = G_N \sum_\sigma (1 - |r_{ee}^\sigma|^2 + |r_{eh}^\sigma|^2), \tag{23}$$

where $G_N = (e^2/h)/(1 + Z^2)$ is the normal-state conductance. Without a barrier between the N and S regions, normal reflections are suppressed and Andreev reflection becomes the only available scattering mechanism. This situation corresponds to taking $Z = 0$ in Eq. (12), such that $|r_{eh}^\sigma|^2 = |v_0/u_0|^2$ and $|r_{ee}^\sigma|^2 = 0$. Consequently, the subgap conductance for transparent junctions, where $|r_{eh}|^2 = 1$, takes the universal value $G_{NS} = 2G_N$ independently of the pairing mechanism; see the green line in Fig. 2(a).

To distinguish between singlet and triplet pairing through conductance measurements, one thus needs junctions that are not fully transmitting and have a strong contribution from normal backscattering of quasi-particles. In Fig. 2(a), we show examples of the conductance for singlet (red) and triplet (blue) superconductors. The latter displays a characteristic zero-bias peak that reveals the presence of a surface Andreev state in the superconductor[18,19,21,22,53,55,56]. In the presence of normal reflections, tunnel spectroscopy of an NIS junction must be done with enough precision to fully resolve the zero-bias peak corresponding to the nontrivial state coming from the $p$-wave superconductor.

The distribution of waiting times between particles, either electrons or holes, at NIS junctions provides complementary information about the pairing mechanism in the superconductor. Figure 2(b) shows the WTDs for reflected holes with the voltage $eV = \Delta/2$. Transparent junctions with $Z \simeq 0$ feature a free flow of Andreev reflected holes, resulting in a WTD both for singlet and triplet superconductors (green line) which is well-approximated by the Wigner-Dyson distribution

$$\mathcal{W}(\tau) = \frac{32\tau^2}{\pi^2 \langle\tau\rangle^3} e^{-4\tau^2/(\pi^2 \langle\tau\rangle^2)}. \tag{24}$$

The Wigner-Dyson distribution is characteristic for the free flow of non-interacting fermions[35,36]. It also reveals several properties of the coherent transport in the NIS junction. First, its maximum is located at the mean waiting time between reflected holes which corresponds to the average current for hole-like excitations, i.e., $\langle \tau_h \rangle = 1/I_h \simeq h/(eV)$. Second, the width of the distribution reveals the wave nature of the quantum excitations - completely regular transport would be characterized by a Dirac delta peak at $\tau = \langle \tau_h \rangle$. Finally, the simultaneous detection of two particles of the same type is forbidden due to the Pauli principle, i.e., $\mathcal{W}_{h \to h}(\tau = 0) = 0$. All these considerations are valid at zero temperature. We do not expect a qualitative change up to finite electronic temperatures comparable to the applied voltage, cf. ref.[36], with the temperature still being smaller than the superconducting gap.

Since normal backscattering is suppressed in transparent junctions, the WTD between reflected electrons approaches Poissonian statistics given by the exponential distribution, $\mathcal{W}_{e \to e}(\tau) \simeq e^{-\tau/\langle\tau\rangle}/\langle\tau\rangle$, as characteristic of low-transmitting contacts[35,36]. Similarly, for a tunnel barrier, Andreev reflections are suppressed for all energies below the gap for singlet superconductor. Therefore, $\mathcal{W}_{h \to h}(\tau)$ also approaches a Poisson distribution; see the red line in Fig. 2(b). On top of that, the WTD features small oscillations with period given by the average particle current [inset of Fig. 2(b)]. In the low transmission regime, the WTD can be expanded in powers of the reflection (or transmission) probability. The oscillations in the WTD are thus well interpreted in terms of the higher order scattering events taking place at the junction[35,36].

The surface state in the triplet superconductor completely changes this picture. For any applied voltage, a perfect Andreev reflection occurs at zero energy, making the scattering probability strongly energy-dependent. The WTD between reflected holes captures this effect resulting in a crossover between Wigner-Dyson and Poisson statistics indicated by the blue line in Fig. 2(b).

In addition to the waiting times between particles of the same type, it is interesting to analyze the distribution of waiting times between different types of particles. These distributions do not necessarily vanish at zero waiting time, since the simultaneous reflection of an electron and a hole is possible. For $p$-wave superconductors, which always fulfill $|r_{eh}(E=0)|^2 = 1$, the distribution of waiting times between electrons and holes is similar to a Wigner-Dyson distribution, but with the important difference that it remains finite at zero waiting time. In Fig. 2(c), we show the distribution of waiting times between electrons and holes for a highly transparent junction, where the conductance cannot clearly resolve the zero-energy state. By contrast, the WTD remains finite at zero waiting time for $p$-wave superconductors (blue line), while it is suppressed to zero for $s$-wave superconductors (red line) and closely follows the Wigner-Dyson distribution (dotted line). Thus, the surface state of the $p$-wave superconductor enables the simultaneous reflection of electrons and holes.

In Fig. 2(d), we analyze in more detail the voltage dependence of the WTDs at short waiting times. For $s$-wave pairing, transport in highly transparent junctions is dominated by Andreev reflections, while normal backscattering is the main microscopic process for tunnel junctions. In both cases, the transport is controlled by only one scattering mechanism, so the WTDs vanish at short waiting times for subgap voltages; cf. red and magenta lines in Fig. 2(d). When the applied voltage is higher than the superconducting gap, both normal and Andreev reflections contribute to the transport through the junction. For transparent junctions, Andreev reflections dominate at subgap energies, but they are suppressed for energies above the gap, where quasi-particle transmission takes place. For tunnel junctions, normal scattering provides the main contribution to transport, except for energies close to the gap, where Andreev processes are enhanced. As a result, we observe a finite WTD at short times for voltages in the range $\Delta \lesssim eV \lesssim 2\Delta$. Junctions with $p$-wave superconductors always feature a finite WTD at short waitings times; see blue and cyan lines in Fig. 2(d). Thus, WTDs can identify topological superconductors with a voltage that is smaller or comparable to the superconducting gap.

## NISIN Junctions

Next, we turn to NISIN junctions. We thus consider a superconductor of finite width and we add a second normal electrode. The resulting structure, with barriers of strength $Z_{L,R}$ on each side of the superconductor, can function as a Cooper pair splitter[63–70]: by biasing the normal leads, Cooper pairs from the superconducting region can be forced to leak into the normal regions. In the ideal case, the two electrons from a split Cooper pair tunnel into different normal leads, while preserving the entanglement of their spins. The time-reversed process, where the incident electrons originate from different electrodes, is known as a crossed Andreev reflection[61,71–82]. If the central region is a topological superconductor, the Majorana edge modes drastically change the transport properties of the system[83–86]. Even for junctions where the superconductor is much larger than the superconducting coherence length, the edge modes allow for a finite conversion of incident electrons from one electrode into transmitted holes in the other. Trivial superconductors, on the other hand, only feature crossed Andreev reflections for energies above the gap.

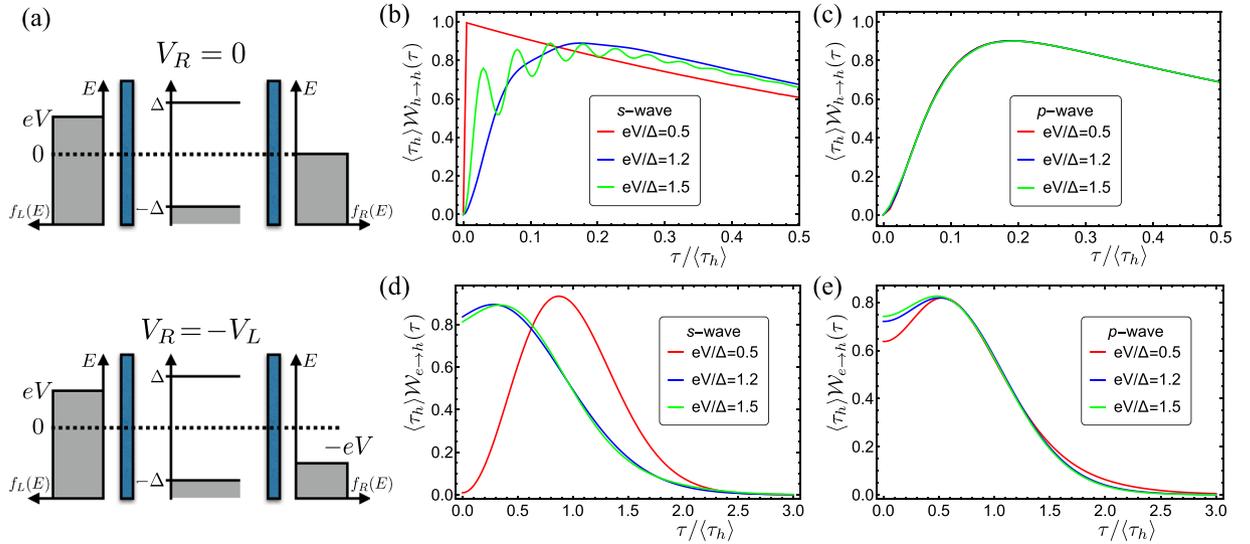

**Figure 3.** Electron waiting times of NISIN junctions. (**a**) Schematics of the setup with two different voltage configurations. In the top panel, a voltage is applied to the left lead, while the right one is kept grounded. In the bottom panel, the leads are biased with opposite voltages. (**b,c**) For $V_R = 0$, the distribution of waiting times between transmitted holes is shown for $s$-wave (**b**) and $p$-wave (**c**) superconductors, respectively. (**d,e**) For $V_R = -V_L$, the distribution of waiting times between transmitted electrons and holes is shown for $s$-wave (**d**) and $p$-wave (**e**) superconductors, respectively. The barrier strengths are $Z_L = 3$ and $Z_R = 2.5$, and $d_S/\xi = 5$ is the width of the superconducting region over the coherence length.

To explore the different transmission profiles for trivial and topological superconductors, we evaluate the WTDs for the NISIN junction. To be specific, we consider low-transmitting barriers and a superconducting region that is wider than the coherence length. As a result, electron and hole transmission through $s$-wave superconductors is very weak and only takes place for energies above the gap. On the other hand, $p$-wave superconductors feature a finite transmission for both electrons and holes at zero energy. However, these scattering processes are of similar magnitude and have opposite contributions to the nonlocal conductance that tend to cancel. Thus, similarly to transparent NIS junctions, the nonlocal conductance cannot clearly identify the presence of topological edge states.

The presence of the edge states can be captured by the WTDs. First, we consider the NISIN junction with a voltage bias applied to the left electrode only, i.e, $V_L = V$ and $V_R = 0$. The WTDs between transmitted hole-like quasi-particles for $s$-wave superconductors strongly depend on the applied voltage, see Fig. 3(b). For subgap voltages (red line), quasi-particle transmission is strongly suppressed and $\mathcal{W}_{h \to h}(\tau)$ approaches an exponential distribution corresponding to Poisson statistics. The two barriers $Z_{L,R}$ create resonance conditions for the transmission of quasi-particles with energies above the gap. If only one of these transmission resonances are located inside the voltage window, the WTD approaches the crossover-regime between Poisson and Wigner-Dyson statistics (blue line). If two or more resonances are located inside the voltage window (green line), the WTD features oscillations due to interference between the different transmission channels[36]. The WTDs for $p$-wave superconductors, by contrast, are very different. Due to the presence of the edge states, $\mathcal{W}_{h \to h}(\tau)$ is dominated by the enhanced transmission at zero energy and follows the crossover-regime between Poisson and Wigner-Dyson statistics, cf. Fig. 3(c). This is the case even for voltages larger than the gap, which include extra quasi-particle transport channels.

We now allow for the voltage to drop symmetrically across the junction by setting $V_L = V$ and $V_R = -V$. In addition to the incoming electrons from the left electrode, we must now include holes being injected from the right lead. These holes experience normal and Andreev reflections at the right interface, which contribute to the outgoing stream of electrons and holes in the right lead. Since we consider low-transmitting interfaces, normal reflections are dominant except for resonant energies. In Fig. 3(d), we show the distribution of waiting times between electrons and holes being transmitted into the right lead from an $s$-wave superconductor. Again, the WTD is very different for voltages above and below the gap. Below the gap, the WTD is well-captured by a Wigner-Dyson distribution and it vanishes at short waiting times. This result is directly connected to the dominant contribution from normal reflection of holes at the right interface. The suppression is lifted for voltages above the gap, since normal transmission of electrons from left to right also has a high probability. The cross-channel WTD for energies over the gap thus has two contributions and follows the WTD for two independent channels described in ref.[36]. These results should be contrasted with those of a $p$-wave superconductor shown in Fig. 3(e). Here, the WTD is nearly independent of the applied voltage, giving a distinct difference from $s$-wave superconductors. As long as the applied voltage is comparable to the superconducting gap, transport in the right lead is dominated by zero-energy scattering processes.

## Conclusions and Outlook

We have shown that waiting time distributions are a useful tool to determine the presence of edge states in topological superconductors. At normal-superconductor junctions, the waiting time distribution of hole-like quasi-particles is equivalent to the distribution of waiting times between Cooper pairs being injected into the superconductor. Moreover, the distribution of waiting times between holes reveals the presence of topological edge states in junctions with $p$-wave superconductors. In addition, the distribution of waiting times between electrons and holes is very sensitive to the edge states, even in transparent junctions, where conductance measurements would not provide a clear signal. In trivial superconductors, the waiting time distribution is suppressed to zero at short waiting times for subgap voltages. By contrast, for topological superconductors, the waiting time distribution remains finite for any applied voltage, providing a clear difference from $s$-wave superconductors. In systems where topological superconductivity is artificially engineered[2–4], the proximity-induced pairing can be a superposition of $s$- and $p$-wave amplitudes. As described in ref.[53], if the $p$-wave amplitude is dominant, a zero energy state emerges. The previous results thus apply to this case.

For NISIN junctions, we have analyzed the waiting time between the transmissions of electrons and holes through the superconducting region. Also in this case, the $p$-wave superconductors feature distinctive behaviors in the nonlocal transport as revealed by the waiting time distributions. For $s$-wave superconductors, the waiting time distribution changes abruptly as the voltage bias is increased above the superconducting gap. By contrast, for $p$-wave superconductors, the transport is dominated by the presence of zero-energy edge modes, and the waiting time distributions are almost independent of the voltage.

As an outlook on future work, we finally point out possible avenues for further developments. In quantum dot systems, the tunneling of individual electrons can now be experimentally observed using single-electron detectors, and measurements of an electron waiting time distribution have recently been reported[30]. In coherent conductors, a measurement of the waiting time distribution seems more challenging, and only recently a quantum theory of an electron waiting time clock has been developed for normal-state conductors[41] with an extension to a spin-sensitive detector being outlined in subsequent work[87]. Recent measurements of the time-of-flight of single-electron excitations through a mesoscopic conductor provide a promising way for directly investigating real-time dynamics of emitted pulses[88,89]. Adapting these ideas to superconducting systems is clearly desirable. Moreover, while we have focused on superconducting junctions with constant voltage biases, it would be interesting to investigate these systems when excited by periodic voltage pulses. The transmission of a charge pulse through a superconducting region may yield additional information about the topological properties of the superconductor, including the presence of edge modes, with clear signatures in the waiting time distribution.

## Acknowledgements

The authors are grateful to M. V. Moskalets for valuable discussions. S.M. and P.B. acknowledge support from the European Union's Horizon 2020 research and innovation program under the Marie Skłodowska-Curie Grants No. 753906 and No. 743884, respectively. This work was performed as part of the Academy of Finland Centre of Excellence program (Project No. 312299).

## Author Contributions

S.M. and P.B. performed calculations. C.F. supervised the research. All authors contributed to the writing of the manuscript.

## Additional Information

**Competing Interests:** The authors declare no competing interests.

**Publisher's note:** Springer Nature remains neutral with regard to jurisdictional claims in published maps and institutional affiliations.